\documentclass[12pt]{article}
\usepackage{amsmath}
\usepackage{graphicx}
\usepackage{enumerate}
\usepackage{natbib}
\usepackage{url} 

\newcommand{\blind}{1}

\numberwithin{equation}{section}
\usepackage{amsfonts}
\usepackage{amssymb}
\usepackage{amsthm}
\newtheorem{theorem}{Theorem}[section]
\newtheorem{lemma}[theorem]{Lemma}

\DeclareMathOperator*{\argmin}{arg\,min}

\begin{document}

\def\spacingset#1{\renewcommand{\baselinestretch}%
{#1}\small\normalsize} \spacingset{1}


\if1\blind
{
  \title{\bf Kernel Estimation of Bivariate Time-varying Coefficient Model for Longitudinal Data With Terminal Event}
  \author{Yue Wang, Bin Nan
  \\
    Department of Statistics, University of California, Irvine\\
    and \\
    Jack D. Kalbfleisch \\
    Department of Biostatistics, University of Michigan}
    \date{}
  \maketitle
} \fi

\if0\blind
{
  \bigskip
  \bigskip
  \bigskip
  \begin{center}
    {\LARGE\bf Kernel Estimation of Bivariate Time-varying Coefficient Model for Longitudinal Data With Terminal Event}
\end{center}
  \medskip
} \fi

\bigskip
\begin{abstract}
We propose a nonparametric bivariate time-varying coefficient model for longitudinal measurements with the occurrence of a terminal event that is subject to right censoring. The time-varying coefficients capture the longitudinal trajectories of covariate effects along with both the followup time and the residual lifetime. The proposed model extends the parametric conditional approach given terminal event time in recent literature, and thus avoids potential model misspecification. We consider a kernel smoothing method for estimating regression coefficients in our model and use cross-validation for bandwidth selection, applying undersmoothing in the final analysis to eliminate the asymptotic bias of the kernel estimator. We show that the kernel estimates are asymptotically  normal under mild regularity conditions, and provide an easily computable sandwich variance estimator. We conduct extensive simulations that show desirable performance of the proposed approach, and apply the method to analyzing the medical cost data for patients with end-stage renal disease.
\end{abstract}

\noindent%
\vfill

\newpage
\spacingset{1.9} 
\section{Introduction}
\label{sec:intro}
In longitudinal studies, it is often the case that the collection of repeated measurements is stopped by the occurrence of some terminal event, for example, death. 
There are two sets of widely used approaches for modelling longitudinal measures with terminal event: the joint modeling approach using latent frailty and the marginal estimating equation approach using inverse probability weighting (IPW).
Under the joint modeling framework, the survival time and the longitudinal process are assumed independent conditional on some latent random effects. Thorough reviews of this type of approach can be found in \cite{tsiatis_joint_2004} and \cite{rizopoulos_joint_2012}. For the marginal estimating equation approach with IPW, readers can refer to \cite{robins_analysis_1995}. These ideas have also been applied to modeling recurrent events in the presence of a terminal event, see e.g., \cite{kalbfleisch_estimating_2013} and \cite{ghosh_marginal_2002}.
They may fall short in certain situations, however.
First, as pointed out by \cite{kong}, they do not explicitly model the association between terminal event time and the longitudinally measured response variable, which is of primary interest in many applications. Second, in health studies where death is a terminal event, some approaches treat the occurrence of death as ``dropout", either informative or non-informative, which implicitly defines the underlying longitudinally measured stochastic processes of health status beyond death. In other words, death causes ``missing data" in such a view, which is questionable since death itself is a fundamental characteristic of health.

For these reasons, several reverse-time models have been considered in the recent literature. \cite{chan} considered a nonparametric approach for the mean of a reverse-time process. \cite{li} considered a likelihood-based approach for the reverse-time model with applications to palliative care, with extension to a semiparametric approach introduced in \cite{li2}. \cite{dempsey} considered reverse alignment as a general technique for constructing models for survival processes and investigated several related statistical consequences. These methods model backward time with event time as the time origin, but lose the interpretation of chronological time effects that are of primary interest in conventional longitudinal studies. To keep the desired chronological time interpretation of regression coefficients and meanwhile to describe the effect of terminal event in longitudinal studies,  \cite{kong} proposed a parametric nonlinear regression model conditional on the terminal event time which builds the residual lifetime into covariate effects. They proposed a two-stage approach that improves the efficiency of parameter estimates of the complete case analysis that only uses data with uncensored event times. But a parametric model can be easily misspecified, and their method cannot handle time-varying covariates that occur overwhelmingly often in longitudinal studies. 


In this article, we propose a nonparametric extension of \cite{kong}. In particular, regression coefficients are bivariate functions of both chronological followup time $t$ and residual lifetime $T-t$ with unknown form, where $t$ denotes the followup time and $T$ denotes the terminal event time. Moreover, time-varying covariates are incorporated in our model. Such a modelling strategy  allows us to assess the varying effect of certain covariate when patients approach death, which is of particular interest for the analysis of end-stage renal disease (ESRD) medical cost data. We estimate the regression coefficients using kernel smoothing and establish the asymptotic normality of kernel estimates together with convergence rate that depends on the bandwidth size. We also provide a consistent sandwich variance estimator that helps construct pointwise confidence bands.


The rest of the article is organized as follows. In Section 2 we introduce the time-varying coefficient model and the kernel estimating method with bandwidth determined via undersmoothing after cross-validation. We outline the asymptotic properties in Section 3 with sketched proofs given in the Appendix. We provide a simulation study in Section 4 and the analysis of ESRD medical cost data in Section 5. We give a few concluding remarks in Section 6.

\section{Modeling Strategy and Estimating Method}

\subsection{Varying Coefficient Model}

Let $Y(t)$ be a stochastic process denoting the response variable measured over time in a longitudinal study. Let $\mathbf{X}(t) = (X_1(t), \dots, X_p(t))$ be $p$ covariate processes. Note that we use bold letter to represent either a vector or a matrix in this article. Suppose the longitudinal cohort data consists of $n$ independent copies of $(Y(t), \mathbf{X}(t))$, representing $n$ individuals' observations in the study cohort, where the $i$th individual's data $(Y_i(t), X_{i1}(t), \dots, X_{ip}(t))$ are measured at random time points $\tau_{ij}$, $j=1, \dots, m_i$. Baseline covariates take constant values over time. 
We define $X_{i1}(t)\equiv1$ for any $i$ and $t$, which determines the intercept. Suppose each individual has $m$ visits, but not all of them are observed because of early stopping due to terminal event or right censoring, which makes the number of actual visits varying among individuals. Specifically for subject $i$, denote the terminal event time as $T_i$ and the right censoring time as $C_i$, then the number of visits of subject $i$ is $m_i=\max\{j: j \le m, \, \tau_{ij}\leq T_i\wedge C_i\}$, where $a \wedge b=\min\{a, b\}$. 
Denote the set of subjects whose terminal events are observed by $\mathcal{D}=\left\{i:T_i \le C_i\right\}$.

We consider the following model for the longitudinal response variable $Y_i$ observed at time $\tau_{ij}$:
\begin{align}
    Y_i(\tau_{ij})=\sum_{k=1}^pX_{ik}(\tau_{ij})\beta_k(\tau_{ij}, T_i-\tau_{ij})+\varepsilon_i(\tau_{ij}),
\label{eq:model}
\end{align}
where each $\varepsilon_i(t)$ is a zero-mean stochastic process with variance function $\sigma^2(t)$ and covariance function $\rho(t_1,t_2)$ for any $t_1\neq t_2$.  Assume all the quantities involved in this model are independent and identically distributed (i.i.d.) across individuals, which include $\{\tau_{ij}\}_{j=1}^m$, $T_i$, $C_i$, $\{X_{ik}(\cdot)\}_{k=1}^p$ and $\varepsilon_i(\cdot)$. Here i.i.d. is defined for processes on any finite index set.
We further assume that for each individual, suppressing the subscript $i$ here without causing any confusion, we have (1) given $\tau_j=t$, $\varepsilon(\tau_j)$ has the same distribution as $\varepsilon(t)$ and is independent of $T$, $C$ and $\{X_{k}(\tau_j)\}_{k=1}^p$; (2) given $(\tau_{j_1}, \tau_{j_2}) = (t_1, t_2)$, $(\varepsilon(\tau_{j_1}), \varepsilon(\tau_{j_2}))$ has the same distribution as $(\varepsilon(t_1), \varepsilon(t_2))$ and is independent of $T$, $C$, $\{X_{k}(\tau_{j_1})\}_{k=1}^p$ and $\{X_{k}(\tau_{j_2})\}_{k=1}^p$. In other words, data observed on a set of random times behave like observed on a set of constant times, which is  commonly assumed in longitudinal data analysis.
Later it will become clear that we do not need to assume $T$ and $C$ are independent or conditionally independent given covariates as one would do in traditional survival analysis.  

Unlike the usual time-varying coefficient model for longitudinal data, a particularly important feature of model (\ref{eq:model}) is that the unknown coefficient $\beta_k(t, T-t)$ is allowed to be a bivariate function not only varying with time since entry, $t$ (the usual setup, see e.g. \citet{hoover}),  but also varying with time from $t$ to the terminal event, $T-t$ (also referred to as residual lifetime if $T$ is death time). Unlike any conventional modeling strategy for longitudinal data with terminal event, allowing $\beta_k$ to depend on $T-t$ directly captures the way in which impending failure modifies the effect of $X_{ik}(t)$. If none of the $\beta_k$, $k=1,\dots,p$, varies with $T-t$, then the above model (\ref{eq:model}) reduces to a standard time-varying coefficient model.      

Model (\ref{eq:model}) extends \cite{kong} from a parametric model to a nonparametric model, from an intercept varying with $T-t$ only to all regression coefficients varying with both $t$ and $T-t$, and from fixed baseline covariates only to time-varying covariates. The model also extends \cite{Lu}, who only considered a nonparametric intercept varying with $T-t$ without pursuing the asymptotic properties of their spline based estimating method. \cite{li_semiparametric_2018} considered a bivariate mean model, included no covariates and did not provide asymptotic results for their spline estimating method.

It becomes clear that model (\ref{eq:model}) is well-defined when $T_i$ is observed, so all the observations collected from time at entry to $T_i$ are complete data, whereas observations collected from time at entry to $C_i$ before $T_i$ are incomplete. This is another major distinction between a model that is conditional on $T_i$ and conventional regression models for longitudinal data with terminal events. Since $T_i$ is subject to right censoring, the problem determined by model (\ref{eq:model}) becomes a regression problem with censored covariate, for which the complete case analysis is a valid approach. This is the method we consider in this article for estimating unknown bivariate coefficient functions $\beta_k$, $k=1,\dots, p$. 
Including observations for censored individuals faces multifaceted difficulties, which will be discussed in Section \ref{sec:disc}.

\subsection{Bivariate Kernel Estimation}

For any fixed point $(t_0, s_0)$, we apply bivariate kernel smoothing to estimate $\boldsymbol{\beta}(t_0, s_0)$ by minimizing the following loss function with respect to $b_k$, $k=1, \dots, p$:
\begin{align}
L_n(t_0, s_0)=\sum_{i\in\mathcal{D}}\sum_{j=1}^{m_i}\left(Y_{ij}-\sum_{k=1}^pX_{ik}(\tau_{ij})b_k\right)^2K\left(\frac{\tau_{ij}-t_0}{h}, \frac{T_i-\tau_{ij}-s_0}{h}\right),
\label{eq:loss}
\end{align}
where $K: \mathbb{R}^2\rightarrow \mathbb{R}$ is the kernel function, $h>0$ is the bandwidth. There are two major distinctions between the resulting estimator from (\ref{eq:loss}) and the estimator in \cite{hoover}: First, the estimator in (\ref{eq:loss}) involves terminal event time $T_i$ and is based on complete data. 
Second, since $\beta_k$'s are bivariate functions, a bivariate kernel function is used. To simplify the numerical implementation, we use the same bandwidth for both time axes and ignore the off-diagonal element of the $2\times 2$ bandwidth matrix.  

Rewrite (\ref{eq:loss}) into the following matrix form:
\begin{align*}
    L_n(t_0,s_0)=\sum_{i\in\mathcal{D}}\left(\mathbf{Y}_i-\mathbf{X}_i\mathbf{b}\right)^T\mathbf{K}_i(t_0,s_0;h)\left(\mathbf{Y}_i-\mathbf{X}_i\mathbf{b}\right), 
\end{align*}
where
\begin{align*}
\mathbf{X}_i=
\begin{pmatrix}
X_{i1}(\tau_{i1}) & \hdots & X_{ip}(\tau_{i1})\\
\vdots & \ddots & \vdots\\
X_{i1}(\tau_{im_i}) & \hdots & X_{ip}(\tau_{im_i})
\end{pmatrix},
\end{align*}
$\mathbf{K}_i(t_0,s_0;h)$ is a diagonal matrix with $j$th element given by $h^{-2} K((\tau_{ij}-t_0)/h, (T_i-\tau_{ij}-s_0)/h)$, and $\mathbf{Y}_i=(Y_{i1},..., Y_{im_i})^T$.
We estimate the time-varying coefficients $\beta_k(t,s)$ by minimizing $L_n(t_0,s_0)$ with respect to $b_k$, $k=1,\dots, p$, i.e.,
\begin{align*}
    \widehat{\boldsymbol{\beta}}(t_0,s_0;h)=\argmin_{\mathbf{b}} L_n(t_0,s_0),
\end{align*}
which has a closed form solution given by
\begin{align}
    \widehat{\boldsymbol{\beta}}(t_0,s_0;h)=\left(\sum_{i\in\mathcal{D}}\mathbf{X}_i^T\mathbf{K}_i(t_0,s_0;h)\mathbf{X}_i\right)^{-1}\left(\sum_{i\in\mathcal{D}}\mathbf{X}_i^T\mathbf{K}_i(t_0,s_0;h)\mathbf{Y}_i\right).
    \label{eq:estimator}
\end{align}
The above estimator given in (\ref{eq:estimator}) ignores the within-subject correlation following the working independence assumption,  which was shown by \cite{lin_nonparametric_2000} to be most efficient when a standard kernel is applied and the cluster size is finite. This  counter-intuitive result was explained by \cite{wang_marginal_2003} who also showed that higher efficiency could be achieved by using an alternative kernel method, which we do not pursue here because of the numerical advantages of the independence assumption and (\ref{eq:estimator}).

\subsection{Automatic Bandwidth Selection and Undersmoothing}

A typical approach for automatic bandwidth selection is through $K$-fold cross-validation (CV). 
To keep the independence between training set and validation set, we partition the longitudinal data at subject level such that all repeatedly measured observations of each subject belong to only one fold. The criterion for selecting bandwidth is to minimize the average predictive squared errors across all validation sets. In particular, let 
$S_k$, $k=1, \dots,K$, be the index set of subjects in the $k$-th fold, where $\cup_kS_k=\mathcal{D}$ and $S_k\cap S_l=\emptyset$ for any $k\neq l$, then the average predictive squared error criterion is given by
\begin{align}
    \textrm{CV}(h)=\frac{1}{\sum_{i\in\mathcal{D}}m_i}\sum_{k=1}^K\sum_{i\in S_k}\sum_{j=1}^{m_i}\left[Y_{ij}-\mathbf{X}_i\widehat{\boldsymbol{\beta}}^{(-k)}(\tau_{ij}, T_i-\tau_{ij};h)\right]^2,
\end{align}
where $\widehat{\boldsymbol{\beta}}^{(-k)}$ represents the kernel estimator calculated by leaving out all observations in the $k$-th fold.
Note that only the complete cases $\mathcal{D}$ are partitioned into folds. In practice, the criterion is minimized on a preselected grid of $h$.


Standard approaches to constructing nonparametric confidence bands for functions are complicated by the impact of bias. According to Hall (1992), bias 
decreases as the amount of statistical smoothing is reduced, which can be clearly seen from the asymptotic distributional results of kernel estimates. Therefore, one way of alleviating bias is to smooth the curve estimator less than would be optimal for point estimation. We choose to undersmooth by multiplying $n^{-\gamma}$ to the selected bandwidth using cross-validation  for some $\gamma>0$. We will see in Section \ref{sec:asymptotics} that undersmoothing still leads to the desirable asymptotic result as long as the undersmoothed bandwidth falls into the range specified by Condition 2 in Appendix \ref{sec:cond}.

\section{Asymptotic Properties}
\label{sec:asymptotics}
\subsection{Asymptotic Normality of $\boldsymbol{\widehat\beta}$}\label{sec:normality}

Under several mild regularity conditions given in the Appendix, we can show that the estimator (\ref{eq:estimator}) follows a multivariate normal distribution as $n$ approaches infinity. First we introduce some notation:
\begin{align}
    \mu_0&=\int K^2(x,y)dxdy,\ 
    \boldsymbol{\mu}_2=
    \begin{pmatrix}
        \int x^2K(x,y)dxdy & \int xyK(x,y)dxdy\\
        \int yxK(x,y)dxdy & \int y^2K(x,y)dxdy
    \end{pmatrix}, \nonumber\\
    \eta(t_0,s_0;j,k,l)&=E\big[1(T\le C)X_k(\tau_j)X_l(\tau_j)|\tau_j=t_0,T=t_0+s_0\big], \nonumber\\
    \boldsymbol{\psi}_k(t_0,s_0)&=\sum_{j=1}^m\sum_{l=1}^p\Big[\nabla\eta(t_0,s_0;j,k,l)f_{\tau_j, T-\tau_j}(t_0,s_0)\nabla\beta_l(t_0,s_0)^T \nonumber\\
    &\quad+\eta(t_0,s_0;j,k,l)\nabla f_{\tau_j, T-\tau_j}(t_0,s_0)\nabla\beta_l(t_0,s_0)^T \nonumber\\
    &\quad+\frac{1}{2}\eta(t_0,s_0;j,k,l)f_{\tau_j, T-\tau_j}(t_0,s_0)\nabla^2\beta_l(t_0,s_0)\Big], \nonumber\\
    \boldsymbol{\Delta}(t_0, s_0)&=h_0^3\Big(\langle\boldsymbol{\mu}_2,\boldsymbol{\psi}_1(t_0,s_0)\rangle, ..., \langle\boldsymbol{\mu}_2,\boldsymbol{\psi}_p(t_0,s_0)\rangle\Big)^T, \nonumber\\
    \mathbf{B}(t_0,s_0)&=\left[\sum^{m}_{j=1}\boldsymbol{\eta}(t_0,s_0;j)f_{\tau_j, T-\tau_j}(t_0,s_0)\right]^{-1}\boldsymbol{\Delta}(t_0, s_0)\label{eq:bias}, \\
    \mathbf{V}(t_0,s_0)&=\left[\sum^{m}_{j=1}\boldsymbol{\eta}(t_0,s_0;j)f_{\tau_j, T-\tau_j}(t_0,s_0)\right]^{-1}\sigma^2(t_0)\mu_0.\label{eq:var}
\end{align}
In the above notation: $\nabla g(t_0, s_0)$ is the gradient of a function $g$ as a column vector at $(t_0,s_0)$ and $\nabla^2g(t_0,s_0)$ is the hessian matrix; $f_{\tau_j, T-\tau_j}$ denotes the joint density of $\tau_j$ and $T-\tau_j$; $h_0$ is the limit of $n^{1/6}h$, which is defined in regularity condition \ref{cond:h_upper}; $\langle A, B\rangle$ is the Frobenius inner product of matrices $A$ and $B$, i.e., $\langle A, B\rangle = \textrm{tr}(A^TB)$; and $\boldsymbol{\eta}(t_0, s_0;j)$ is the $p\times p$ matrix with the $(k,l)$-th element given by $\eta(t_0, s_0; j,k,l)$. Also note that the subscript $i$ is suppressed in all above quantities to simplify the notation since these are all defined for a generic subject and observations are assumed i.i.d. 

\begin{theorem}\label{thm:normal}
    Under regularity conditions \ref{cond:K}-\ref{cond:pd} in Appendix \ref{sec:cond}, with $t_0, s_0>0$, the following asymptotic normality holds for $\widehat{\boldsymbol{\beta}}(t_0,s_0;h)$ given in (\ref{eq:estimator}):
    \begin{align}
        n^{1/2}h\left(\widehat{\boldsymbol{\beta}}(t_0,s_0;h)-\boldsymbol{\beta}(t_0,s_0)\right)\to_dN\left(\mathbf{B}(t_0,s_0),\mathbf{V}(t_0,s_0)\right)
    \end{align}
    as $n\to\infty$.
\end{theorem}

Directly using formulae (\ref{eq:bias})-(\ref{eq:var}) to estimate the asymptotic bias and variance is difficult due to their complicated forms. 
We will consider a simpler and numerically implementable sandwich estimator for the asymptotic variance in the following Section \ref{sec:sandwith}, which is shown to be consistent. 
We avoid estimating the bias via undersmoothing. Specifically, we eliminate the asymptotic bias by shrinking the selected bandwidth under cross-validation by a factor of $n^{-1/20}$,
which shows satisfactory performance in simulations.

\subsection{Sandwich Variance Estimator and Pointwise Confidence Band}
\label{sec:sandwith}
With undersmoothing, the asymptotic bias in Theorem \ref{thm:normal} disappears when $n$ goes to infinity. Hence we only need to  estimate the variance $\mathbf{V}(t_0,s_0)$ in order to construct the confidence bound. It turns out that the following sandwich estimator is a valid variance estimator:
\begin{align}
    \widehat{\mathbf{V}}(t_0,s_0)=nh^2\left(\sum_{i\in\mathcal{D}}\mathbf{X}_i^T\mathbf{K}_i\mathbf{X}_i\right)^{-1}\left(\sum_{i\in\mathcal{D}}\mathbf{X}_i^T\mathbf{K}_i\widehat{\boldsymbol{\varepsilon}}_i\widehat{\boldsymbol{\varepsilon}}_i^T\mathbf{K}_i\mathbf{X}_i\right)\left(\sum_{i\in\mathcal{D}}\mathbf{X}_i^T\mathbf{K}_i\mathbf{X}_i\right)^{-1},
    \label{eq:variance}
\end{align}
where $\widehat{\boldsymbol{\varepsilon}}_i$ is the residual vector for the $i$-th subject and $\mathbf{K}_i$ is short for $\mathbf{K}_i(t_0, s_0;h)$. The elements of $\widehat{\boldsymbol{\varepsilon}}_i$ are calculated by
\begin{align*}
    \widehat{\varepsilon}_{ij} =\widehat{\varepsilon}_i(\tau_{ij})  =Y_i(\tau_{ij})-\sum_{k=1}^pX_{ik}(\tau_{ij})\widehat{\beta}_k(\tau_{ij},T_i-\tau_{ij}),\quad 1\leq j\leq m_i.
\end{align*}
The following theorem demonstrates the consistency of (\ref{eq:variance}).
\begin{theorem}\label{thm:consistency}
    Under assumptions of Theorem \ref{thm:normal}, we have
    \begin{align*}
        \widehat{\mathbf{V}}(t_0,s_0)\to_p\mathbf{V}(t_0,s_0)
    \end{align*}
    as $n\to \infty$.
\end{theorem}
With this theorem, an approximate $1-\alpha$ pointwise confidence interval of $\beta_k(t_0, s_0)$ without bias correction can be constructed as
\begin{align*}
    \widehat{\beta}_k(t_0, s_0) \pm z_{\alpha/2}\left(\frac{1}{nh^2}\widehat{\mathbf{V}}(t_0,s_0)_{kk}\right)^{1/2}.
\end{align*}

\section{A Simulation Study}
This section reports the numerical performance of the kernel estimator (\ref{eq:estimator}). For the simulation study, consider model (\ref{eq:model}) with $p=3$ and $m=20$, where the coefficients are the following functions:
\begin{align*}
    &\beta_1(x,y)=\frac{x}{4}\exp\left(-\frac{x^2+y^2}{100}\right),  
    &\beta_2(x,y)=\frac{1}{2}\left[\sin\left(\frac{2x}{5}\right)-\sin\left(\frac{y}{2}\right)\right], 
    &\beta_3(x,y)=\cos\left(\frac{x^2+y^2}{100}\right).
\end{align*}
These functions are similar to those used in \cite{wu}. We generate visiting times in the following way: for subject $i$, the first visit time $\tau_{i1}$ is generated uniformly on $[0,1]$, then $\tau_{ij}$, $j>1$, is generated independently from $\tau_{ij}-(j-1)\sim\textrm{Beta}(\tau_{i1}/4\nu^2, (1-\tau_{i1})/4\nu^2)$, where $\nu$ serves as an upper bound of standard deviation of the Beta distribution and is set to be 0.01. The generated interarrival time $\tau_{ij}-\tau_{i,j-1}$ falls into $[0,2]$ with mean 1 and a very small variance. Thus the generated visiting schedule is approximately evenly spaced, mimicking a designed longitudinal study with annual visits. For covariates, $X_{1}$ is always 1, $X_{2}$ is generated from a standard normal distribution, and $X_{3}(t)$ is a mean-zero Gaussian process with covariance function $Cov\left(X_{3}(t),X_{3}(s)\right)=\exp(-(t-s)^2)$. Moreover, $X_{2}$ and $X_{3}(t)$ are correlated with covariance $Cov\left(X_{2}, X_{3}(t)\right)=0.8\exp(-t^2)$.  Terminal event time $T$ and right censoring time $C$ are generated from exponential distributions with intensities $\exp(3X_{i2}+X_{i3}(0)-5)$ and $\exp(X_{i2}+3X_{i3}(0)-5)$, respectively, and conditionally independent given covariates for convenience. In fact, we do not require conditional independence of $T$ and $C$ for the proposed complete case analysis. To avoid 
the situation that too many visits are censored, we truncate both exponential distributions at 15 and then add a constant 5. 
This yields about 50\% censoring rate. The error term $\varepsilon_i(\tau_{ij})$ is generated by a nonhomogeneous Ornstein–Uhlenbeck (NOU) process $U_i(t)$ plus a random error. The NOU process satisfies $Var(U_i(t))=\exp(1-0.1t)$ and $Corr(U_i(t_1),U_i(t_2))=0.5^{|t_1-t_2|}$, and the random error follows a standard normal distribution.

\begin{figure}[h!]
    \centering
    \includegraphics[scale=0.5]{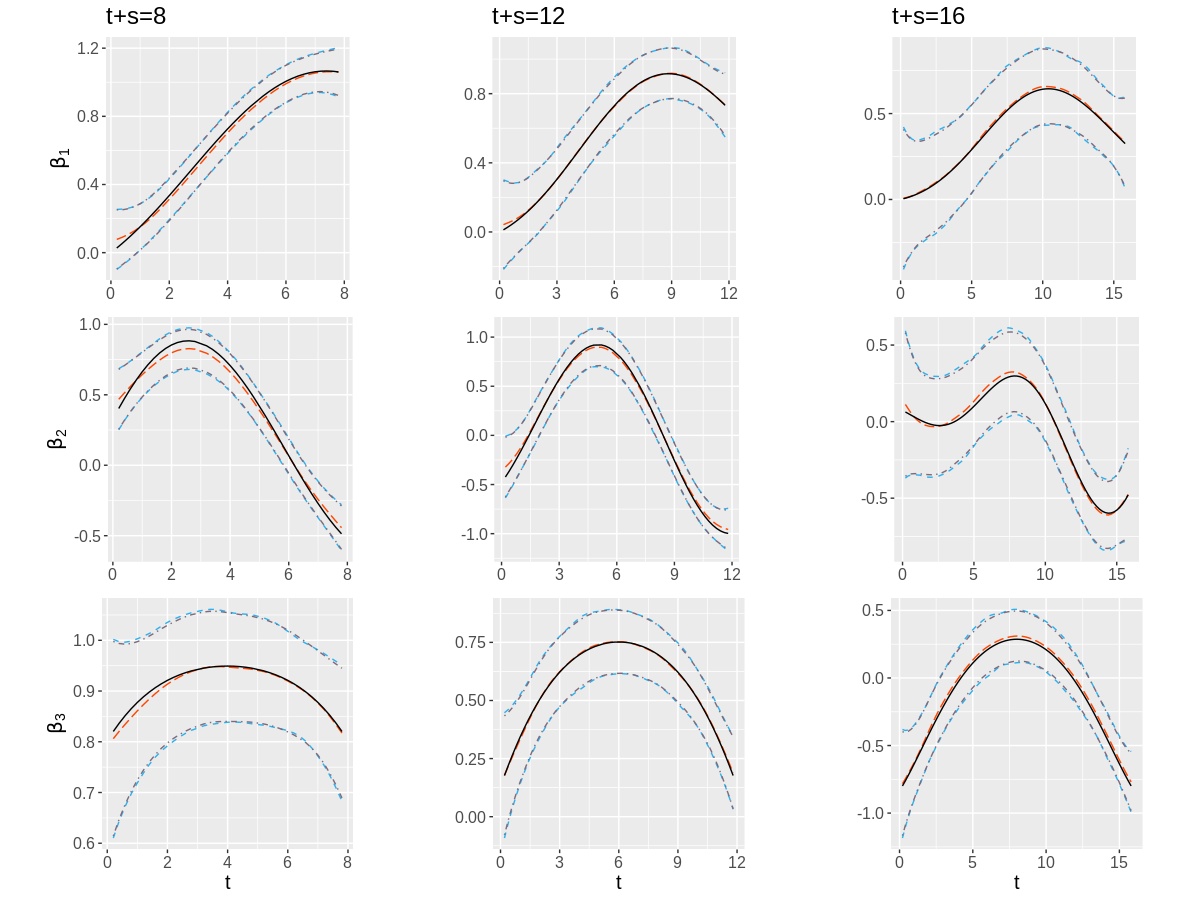}
    \caption{Regression coefficients and their point-wise confindence bands at $T=8$, $12$, and $16$.}
    \label{fig:simulation}
\end{figure}

With this design, we simulate 1000 independent replications, each with a sample size $n=4000$. The kernel function is the density of a standard bivariate normal distribution truncated by a circle around $(0,0)$ which contains probability 0.95. The undersmoothing factor is set at $n^{-1/20}\approx 0.6605$.
To achieve a better visual effect of bivariate functions, we plot a few slices of estimated coefficients. Specifically, we plot $\widehat\beta_k(t, T-t)$ varying with $t$ at $T=8$, $12$, and $16$, separately, which are the estimated covariate effects  from the time of entry to the terminal event for individuals who died at time 8, 12 and 16. 
Among the $3\times3$ panels in Figure \ref{fig:simulation}, each row represents a time-varying coefficient and each column represents one chosen value of $T$. There are 6 curves in each panel: the true function $\beta_k$ (solid), the estimator $\widehat{\beta}_k$ (long dashed), upper and lower confidence bounds calculated from the empirical standard error of 1000 estimates (dashed), and upper and lower confidence bounds calculated from the mean of 1000 standard error estimates (dot-dashed). 
We can see that across all panels, the undersmoothing yields negligible biases, and the two types of confidence bounds are well overlapped, indicating the validity of the proposed variance estimator.



To have an overall view of the performance of the 95\% point-wise confidence bands, we further provide a heatmap of their coverage probability over the entire support of both time axes in Figure 2. It is clear to see that the coverage probability is around 95\% in most area, but can drop to near 80\% on the boundaries or regions where the curvature of the coefficient is large due to relatively large biases of kernel smoothing in such regions. 

\section{The ESRD Medicare Data Analysis}
We consider inpatient medical cost data of patients with end-stage renal disease (ESRD) from year 2007 to 2018 collected by the United States Renal Data System (USRDS). The longitudinal response is the daily inpatient cost paid by Medicare and the terminal event is death. 
The pattern of end-of-life Medicare cost has been identified in previous work. 
For example, \cite{chan} showed an increasing and then decreasing pattern in Medicare costs before death among ovarian cancer patients.
\cite{liu} found an increasing pattern in monthly outpatient EPO costs starting from 6 months prior to death and an initial jump since entry time, followed by a linear drop. When it comes to inpatient cost among ESRD patients, \cite{kong} established similar initial and terminal patterns using a parametric model. Here we aim to verify the patterns using our nonparametric modeling approach.

Following \cite{kong}, we only include black and white patients who started their ESRD services in 2007 and were at least 65 years old when they started. Other than that, we exclude patients who received kidney transplant because they could potentially have very different trajectories of inpatient costs. 
Instead of selecting a simple random sample of available ESRD patients for the analysis as in \cite{kong}, all eligible patients are included in our analysis. Additionally, we are able to take advantage of the most updated data from USRDS, for which the follow-up ended on June 30th, 2018. We end up with a much larger sample size of 42,253 patients who died before the end of follow-up, much longer follow-up with an average of 34.6 months, and a very low censoring rate of only 3.74\%.  In the original data, a total cost  is given for each hospitalization period. To convert the total cost into daily cost, we assume the cost rate is constant during each hospitalization. For example, if \$1000 is claimed for a patient during one hospitalization from May 1st to 20th, then the daily cost on May 10th is calculated to be \$50. If the subject is not in hospital on a certain date, the cost is zero. We choose to select a 5\% random sample of all days from entry to end of observation for each subject and calculate daily costs for these days. This greatly shortens the computation time for cross-validation and is valid because we do not require independence between observation time $\tau_j$ and time to death $T$.

We have considered the same set of covariates as in \cite{kong}, which includes race, heart disease and diabetes. Additionally, we include a binary covariate that indicates if Medicare is a secondary payer. Although most ESRD patients are eligible to apply for Medicare as their primary insurance payer, some are not immediately eligible for Medicare primary payer coverage at retirement because of their employment status and pre-existing primary insurance payers (e.g., group health plans). For this reason, this indicator is time-varying in the follow-up period. Since the effect sizes of the original set of covariates in \cite{kong} are very close to zero and their 95\% confidence bands contain zero over almost the entire support of $(\tau_{ij}, T_i-\tau_{ij})$, due to lack of space, we only display the results for the following simple model for the population-level cost trajectory with a single binary time-varying covariate: 
\begin{align}\label{eq:usrds}
    \log(Y_{ij}/1000+1)&=\beta_1(\tau_{ij}, T_i-\tau_{ij})+\beta_2(\tau_{ij}, T_i-\tau_{ij})1\{\textrm{Medicare is secondary payer at } \tau_{ij}\}\nonumber\\
    &\quad+\varepsilon_i(\tau_{ij}).
\end{align}
We have found that the estimated parameters in the above model are almost identical to those obtained with the additional baseline covariates.

In Figure \ref{fig:usrds}, we plot the estimated curves and their confidence bands for $\beta_1$ and $\beta_1+\beta_2$, respectively, where $\beta_1$ represents the log-transformed daily Medicare payment trajectory among ESRD patients when Medicare is the primary payer, and $\beta_1+\beta_2$ corresponds to Medicare the secondary payer. Similar to how we display simulation results, we choose to only visualize $\beta_k(t, T-t)$ under several fixed values of $T$. Here we choose $T=360$, $900$, and $1440$, corresponding to patients who died roughly 1 year, 2.5 years and 4 years after entry, respectively.

From Figure \ref{fig:usrds} we see that the Medicare cost as primary payer starts to escalate from roughly 150 days prior to death,  similar to the pattern observed in \cite{kong}. The peak value, however, is at the time of death, which is different to \cite{kong} where the peak was around three weeks before death. The initial pattern is different too: in our analysis, we find that the Medicare cost decreases drastically in the first two months after entry, then becomes stabilized overtime until close to death, whereas in \cite{kong}, and in \cite{liu} as well, it increases first then decreases. Such differences are likely due to the restrictive parametric assumptions imposed in \cite{kong} and \cite{liu}.  We also find that, when Medicare is secondary payer, the pattern of inpatient costs is similar but the magnitude is much smaller and at most times very close to zero. This is anticipated because a large portion of medical costs was paid first by some other insurance.



\section{Discussion}
\label{sec:disc}
We propose a nonparametric bivariate time-varying coefficient model for longitudinal data and a kernel estimating method based on complete cases analysis, i.e. using data when terminal events are available, which is shown to be a valid approach. In contrast, \cite{kong} considered a parametric model and likelihood based estimating method that uses all observations, including both censored and uncensored observations. Clearly parametric models suffer from model misspecification, but \cite{kong} showed that including censored observations improves efficiency over the complete case analysis. We find the likelihood based inference approach is difficult to apply for models involving time-varying covariates for two reasons: (1) It requires estimating the survival function $P(T>t|\widetilde{X}(t))$ beyond censoring time, i.e. $t>C$, if one wants to include censored individuals, where $\widetilde{X}(t)$ is the history of the time-dependent covariates $X$ up to time $t$. This means that the covariate history has to be extrapolated beyond $C$ as \cite{kong} also pointed out, which is extremely difficult and introduces measurement errors even when it is doable. (2) Oftentimes in longitudinal studies, the time-varying covariates are internal covariates \citep{kalbfleisch_prentice2002} that make the conditional survival function undefined, leading to invalid likelihood based inference. Hence we argue that the complete case analysis is most appropriate for longitudinal data with time-varying covariates, as in our analysis of the USRDS data where the indicator variable of Medicare as secondary payer is time-varying. Additionally, efficiency loss should not be a concern in our analysis of the USRDS data because of the very low censoring rate. It is worth pointing out that even under scenarios where efficiency loss could be a serious concern, our proposed method can be used as an exploratory tool for finding an appropriate parametric model.


The kernel estimator (2.3) and its asymptotic properties are motivated by \cite{wu} with extensions to bivariate time-varying coefficients. However, there are major differences between our setup and theirs. First, we assume fixed maximum number of observations $m$ instead of letting $m$ go to infinity as the sample size $n$ approaches infinity. This is categorized as the sparse functional case by \cite{Hsing} and induces a simplified version of the asymptotic variance, where the correlation between observations vanishes. Second, each observation time $\tau_j, 1\leq j\leq m$, is allowed to have its own distribution rather than being i.i.d. As a result, $\tau_j$ can flexibly depend on the history up to $\tau_{j-1}$, which reflects more practical settings of longitudinal studies.

\appendix


\section{Appendix}
\subsection{Figures}
\begin{figure}[h!]
    \centering
    \vspace{-0.2in}
    \includegraphics[scale=0.7]{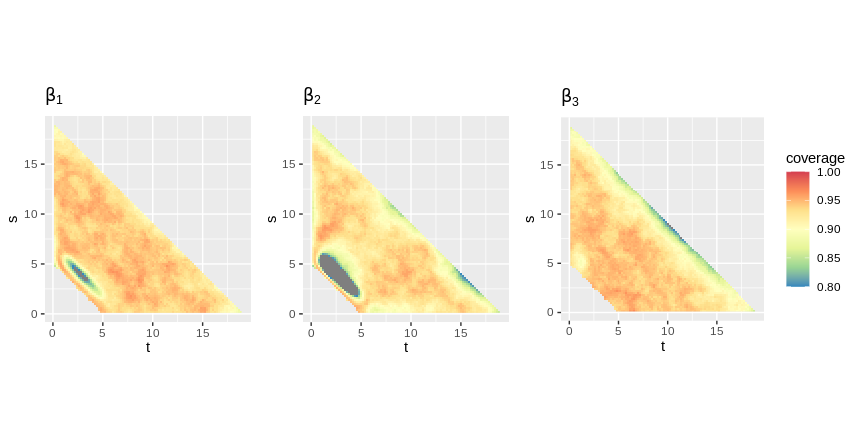}
    \vspace{-0.5in}
    \caption{Coverage probabilities of 95\% pointwise confidence intervals for $\beta_1$, $\beta_2$ and $\beta_3$.}
\end{figure}
\begin{figure}[h!]
    \centering
    \includegraphics[scale=0.5]{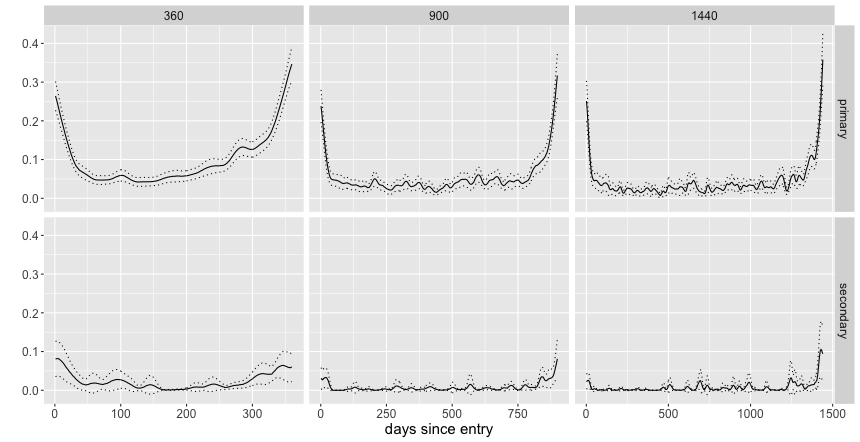}
    \caption{Estimated curves and their confidence bounds for $\beta_1$ (Medicare as primary payer) and $\beta_1+\beta_2$ (Medicare as secondary payer) under $T=360, 900, 1440$.}
    \label{fig:usrds}
\end{figure}

\subsection{Regularity Conditions} 
\label{sec:cond}
Denote $C^d$  the class of functions with $d$-th order continuous derivatives. We introduce a set of regularity conditions:
\begin{enumerate}
        \item $K$ is a probability density of the form $f(\|\mathbf{x}\|_2)$, where $f(\cdot)$ is of bounded variation on bounded support. \label{cond:K}
    \item \begin{enumerate}
    \item $n^{1/6}h\to h_0<\infty$, where $h_0\geq0$. \label{cond:h_upper}
    \item $n^{3/4}h^2/\log n\to\infty$.
    \label{cond:h_lower}
    \end{enumerate} 
    \item \begin{enumerate}
        \item For any $j,k,l$, $\eta(x,y;j,k,l)=E[1(T\le C)X_k(\tau_j)X_l(\tau_j)|\tau_j=x,T=x+y]$ is of class $C^1$ in a neighborhood of $(x,y)=(t_0,s_0)$. \label{cond:X2_cont}
        \item For any $j,k$, $E\left[X_k(\tau_j)^8|\tau_j=x, T=x+y\right]$ is bounded in a neighborhood of $(x,y)=(t_0, s_0)$. \label{cond:X8_bounded}
        \item For any $j_1\neq j_2$ and $k$, $E\left[X_k(\tau_{j_1})^8|\tau_{j_1}=x, \tau_{j_2}=y, T=x+z\right]$ is bounded in a neighborhood of $(x,y,z)=(t_0, t_0, s_0)$. \label{cond:X83_bounded}
    \end{enumerate}
    \item \begin{enumerate}
        \item For any $j$, $f_{\tau_j, T-\tau_j}$ is of class $C^1$ in a neighborhood of $(t_0,s_0)$. \label{cond:f_cont}
        \item For $j_1\neq j_2$, $f_{\tau_{j_1}, \tau_{j_2}, T}$ is bounded in a neighborhood of $(t_0, t_0, t_0+s_0)$.\label{cond:f3_bounded}
    \end{enumerate}
    \item For any $k$, $\beta_k(t,s)$ is of class $C^2$ in a neighborhood of $(t_0,s_0)$. \label{cond:beta_cont}
    \item \begin{enumerate}
        \item $\sigma^2(t)$ is continuous at $t_0$. \label{cond:sigma_cont}
        \item $E\varepsilon(t)^4$ is bounded in a neighborhood of $t_0$. \label{cond:sigma4_bounded}
    \end{enumerate}
    \item There exists $j$ such that $\boldsymbol{\eta}(t_0,s_0;j)$ is positive definite and $f_{\tau_j,T-\tau_j}(t_0,s_0)$ is positive. \label{cond:pd}
\end{enumerate}

Remark: Most of the regularity conditions are direct extensions of those in \cite{wu} to the bivariate case. Specifically, Condition \ref{cond:K} ensures that $K$ has a compact support on $\mathbb{R}^2$ and is symmetric, i.e.,
\begin{align*}
    \int\!\!\!\int xK(x, y)dxdy=0,\quad \int\!\!\!\int yK(x, y)dxdy=0.
\end{align*}
Conditions \ref{cond:h_upper} and \ref{cond:h_lower} together specify a range of feasible bandwidths, which justifies the use of undersmoothed bandwidth. The assumptions of finite  higher order moments for $X$ and $\varepsilon$ in Conditions \ref{cond:X8_bounded}, \ref{cond:X83_bounded} and \ref{cond:sigma4_bounded}  are needed for the consistency of our proposed sandwich variance estimator, which automatically holds for sub-Gaussian or sub-exponential processes. Lastly, Condition \ref{cond:pd} ensures that $\sum_{i\in\mathcal{D}}\mathbf{X}_i^T\mathbf{K}_i\mathbf{X}_i/n$ is invertible asymptotically. This is commonly assumed for regression models.

\subsection{Technical Lemmas}
\begin{lemma}
    For any fixed $n$, let $X_{n,i}$,  $i=1,...,n$, be i.i.d random variables with mean $\mu_n$ and variance $\sigma^2_n$, where $\sigma^2_n=o(n)$. If $\mu_n=O(1)$, then $\sum^n_{i=1}X_{n,i}/n=O_p(1)$; if $\mu_n\to\mu$, then $\sum^n_{i=1}X_{n,i}/n\to_p\mu$.
    \label{lem:lln}
\end{lemma}
\begin{lemma}
    \label{lem:four_parts}
    Let $A_n(\mathbf{x})$ and $A(\mathbf{x})$ be symmetric matrix-valued functions,  and $B_n(\mathbf{x})$ and $B(\mathbf{x})$ be matrix-valued functions of $\mathbf{x}\in\mathcal{X}$, where $B_n(\mathbf{x})$ and $B(\mathbf{x})$ may not be square matrices and can potentially be vectors. Let $\|\cdot\|_2$ denote the spectral norm, i.e., $\|A\|_2=\sup_{\mathbf{x}\neq0}\|A\mathbf{x}\|_2/\|\mathbf{x}\|_2$, and $\lambda_1\big(A(\mathbf{x})\big)$ be the smallest eigenvalue of $A(\mathbf{x})$. Suppose the following hold for a sequence $C_1,C_2,...$ of subsets of $\mathcal{X}$: 
    \begin{align*}
        &\sup_{\mathbf{x}\in C_n}\|A_n(\mathbf{x})-A(\mathbf{x})\|_2\to_p0,\\
        &\sup_{\mathbf{x}\in C_n}\|B_n(\mathbf{x})-B(\mathbf{x})\|_2\to_p0,\\
        &\lim\inf_n\inf_{\mathbf{x}\in C_n}\lambda_1\big(A(\mathbf{x})\big)=a>0,\\
        &\lim\sup_n\sup_{\mathbf{x}\in C_n}\|B(\mathbf{x})\|_2=b<\infty.
    \end{align*}
     Then we have
    \begin{align*}
        \sup_{\mathbf{x}\in C_n}\|A_n(\mathbf{x})^{-1}B_n(\mathbf{x})-A(\mathbf{x})^{-1}B(\mathbf{x})\|_2\to_p0.
    \end{align*}
\end{lemma}

\begin{lemma}
    If the class of functions $\mathcal{F}$ is Euclidean \citep{nolan_u-processes_1987} for a constant envelope function $F$, $0<F<\infty$, and $g$ is a fixed  function, then the class
    \begin{align*}
        \mathcal{F}g=\{fg:f\in\mathcal{F}\}
    \end{align*}
    is Euclidean for the envelope function $F|g|$.
    \label{lem:fg_euclidean}
\end{lemma}

\subsection{Sketched Proofs of Theorem \ref{thm:normal} and \ref{thm:consistency}}
Here we sketch the proofs of Theorems \ref{thm:normal} and \ref{thm:consistency}.
\subsubsection{Sketched Proof of Theorem \ref{thm:normal}: Asymptotic Normality}
\label{sec:proof1}
Define $R_n(t_0,s_0;h) = B_n(t_0,s_0;h)-A_n(t_0,s_0;h)\boldsymbol{\beta}(t_0,s_0)$, where
\begin{align*}
    &A_n(t_0,s_0;h)_{kl} =\frac{1}{n}\sum_{i\in\mathcal{D}}\mathbf{X}_{ik}^T\mathbf{K}_i(t_0,s_0;h)\mathbf{X}_{il},\\
    &B_n(t_0,s_0;h)_{k} =\frac{1}{n}\sum_{i\in\mathcal{D}}\mathbf{X}_{ik}^T\mathbf{K}_i(t_0,s_0;h)\mathbf{Y}_i.
\end{align*}
We can see that $\widehat{\boldsymbol{\beta}}(t_0, s_0)-\boldsymbol{\beta}(t_0,s_0)=A_n(t_0,s_0;h)^{-1}R_n(t_0,s_0;h)$. The proof proceeds by showing that $A_n(t_0,s_0;h)$ converges to a fixed matrix and $\sqrt{nh^2}R_n(t_0,s_0;h)$ converges to a normal random vector.

To show the consistency of $A_n(t_0,s_0;h)$, we apply Lemma \ref{lem:lln} by examining its expectation and variance. With some simple algebra and applying the dominated convergence theorem, we can show
\begin{align*}
    &EA_n(t_0,s_0;h)_{kl}=\sum^{m}_{j=1}\eta(t_0,s_0;j,k,l)f_{\tau_j,T-\tau_j}(t_0,s_0) + o(1),\\
    &Var\left(A_n(t_0,s_0;h)_{kl}\right)=O\left(\frac{1}{nh^2}\right).
\end{align*}
Then by Lemma \ref{lem:lln}, we have
\begin{align}
    A_n(t_0,s_0;h)\to_p\sum^{m}_{j=1}\boldsymbol{\eta}(t_0,s_0;j)f_{\tau_j,T-\tau_j}(t_0,s_0).
    \label{eq:lln_A}
\end{align}

For random vector $R_n(t_0,s_0;h)$, we apply the Cramer-Wold device by showing the asymptotic normality of $\sqrt{nh^2}\boldsymbol{a}^TR_n(t_0,s_0;h)$ for an arbitrary non-zero constant vector $\boldsymbol{a}$. We adopt the Lyapunov's central limit theorem by verifying the following  Lyapunov's condition:
\begin{align*}
    \frac{1}{s_n^{2+\delta}}\sum^n_{i=1}E\left|Z_{n,i}(\boldsymbol{a})-EZ_{n,i}(\boldsymbol{a})\right|^{2+\delta}\to0
\end{align*}
for some $\delta>0$, where $Z_{n,i}(\boldsymbol{a})=1(i\in\mathcal{D})\boldsymbol{a}^T\mathbf{X}_i^T\mathbf{K}_i(t_0,s_0;h)\left[\mathbf{Y}_i-\mathbf{X}_i\boldsymbol{\beta}(t_0,s_0)\right]$ and $s_n^2 = \sum^n_{i=1}\textrm{Var}(Z_{n,i}(\boldsymbol{a}))$. We examine the expectation,  variance, and $(2+\delta)$-th central moment of $Z_{n,i}(\boldsymbol{a})$, and obtain the following:
\begin{align*}
    &EZ_{n,i}(\boldsymbol{a})=h^2\sum_{k=1}^pa_k\langle\boldsymbol{\mu}_2, \boldsymbol{\psi}_k(t_0,s_0)\rangle+o(h^2),\\
    &Var(Z_{n,i}(\boldsymbol{a}))=\frac{1}{h^2}\sum_{j=1}^m\boldsymbol{a}^T\boldsymbol{\eta}(t_0,s_0;j)\boldsymbol{a}\cdot f_{\tau_j,T-\tau_j}(t_0,s_0)\sigma^2(t_0)\mu_0+o\left(\frac{1}{h^2}\right),\\
    &E|Z_{n,i}(\boldsymbol{a}) -EZ_{n,i}(\boldsymbol{\alpha})|^{2+\delta}=O\left(h^{-2-2\delta}\right).
\end{align*}
Thus we have $s_n^2 \ge C n/h^2$ for some constant $C >0$, and with large enough $n$,
\begin{align*}
     &\frac{1}{s_n^{2+\delta}}\sum^n_{i=1}E\left|Z_{n,i}(\boldsymbol{a})-EZ_{n,i}(\boldsymbol{a})\right|^{2+\delta}\leq C'\frac{n/h^{2+2\delta}}{n^{1+\delta/2}/h^{2+\delta}}=C'n^{-\delta/2}h^{-\delta}\to0 
\end{align*}
by Condition \ref{cond:h_lower}. Now we can claim
\begin{align*}
    \frac{\sum_iZ_{n,i}(\boldsymbol{a})-\sum_iEZ_{n,i}(\boldsymbol{a})}{\sqrt{\sum_{i=1}^nVar\left(Z_{n,i}(\boldsymbol{a})\right)}}\to_d N(0,1).
\end{align*}
Then by the Cramer-Wold device, we have
\begin{align*}
    \sqrt{nh^2}R_n(t_0,s_0;h)\to_dN\bigg(\boldsymbol{\Delta}(t_0,s_0),
    \boldsymbol{\Sigma}(t_0, s_0)\bigg),
\end{align*}
where 
\begin{align*}
    &\boldsymbol{\Delta}(t_0,s_0)=h_0^3
    \begin{pmatrix}
        \langle\boldsymbol{\mu}_2,\boldsymbol{\psi}_1(t_0,s_0)\rangle\\
        \vdots\\
        \langle\boldsymbol{\mu}_2,\boldsymbol{\psi}_p(t_0,s_0)\rangle
    \end{pmatrix},\\
    &\boldsymbol{\Sigma}(t_0, s_0)=\sum_{j=1}^m\boldsymbol{\eta}(t_0,s_0;j)f_{\tau_j,T-\tau_j}(t_0,s_0)\sigma^2(t_0)\mu_0.
\end{align*}
Further by (\ref{eq:lln_A}) and the Slutsky's theorem, we obtain Theorem \ref{thm:normal}.

\subsubsection{Sketched Proof of Theorem \ref{thm:consistency}: Consistency of the Sandwich Variance Estimator}
Since we already have obtained  (\ref{eq:lln_A}), now we only need to show
\begin{align*}
    \frac{h^2}{n}\sum_{i\in\mathcal{D}}\mathbf{X}_i^T\mathbf{K}_i\widehat{\boldsymbol{\varepsilon}}_i\widehat{\boldsymbol{\varepsilon}}_i^T\mathbf{K}_i\mathbf{X}_i\to_p\sum_{j=1}^{m}\boldsymbol{\eta}(t_0,s_0;j)f_{\tau_j,T-\tau_j}(t_0,s_0)\sigma^2(t_0)\mu_0.
\end{align*}
Note that
\begin{align*}
    &\left[\frac{h^2}{n}\sum_{i\in\mathcal{D}}\mathbf{X}_i^T\mathbf{K}_i\widehat{\boldsymbol{\varepsilon}}_i\widehat{\boldsymbol{\varepsilon}}_i^T\mathbf{K}_i\mathbf{X}_i\right]_{kl}\\
    & \qquad =\frac{h^2}{n}\sum^n_{i=1}\sum_{j_1=1}^m\sum_{j_2=1}^m1(\tau_{ij_1}\vee \tau_{ij_2}\leq T_i\le C_i)X_{ik}(\tau_{ij_1})X_{il}(\tau_{ij_2})K_{ij_1}K_{ij_2}\\
    &\qquad \quad\times \left\{X_i(\tau_{ij_1})^T\left[\boldsymbol{\beta}(\tau_{ij_1}, T_i-\tau_{ij_1})-\widehat{\boldsymbol{\beta}}(\tau_{ij_1}, T_i-\tau_{ij_1})\right]+\varepsilon_i(\tau_{ij_1})\right\}\\
    &\qquad \quad\times \left\{X_i(\tau_{ij_2})^T\left[\boldsymbol{\beta}(\tau_{ij_2}, T_i-\tau_{ij_2})-\widehat{\boldsymbol{\beta}}(\tau_{ij_2}, T_i-\tau_{ij_2})\right]+\varepsilon_i(\tau_{ij_2})\right\},
\end{align*}
where $K_{ij} = h^{-2}K((\tau_{ij}-t_0)/h, (T_i-\tau_{ij}-s_0)/h)$. By similar mean and variance calculations to those in Section \ref{sec:proof1} following Lemma \ref{lem:lln}, we have
\begin{align*}
    &\frac{h^2}{n}\sum^n_{i=1}\sum_{j_1=1}^m\sum_{j_2=1}^m|X_{ik}(\tau_{ij_1})X_{il}(\tau_{ij_2})|\|X_i(\tau_{ij_1})X_i(\tau_{ij_2})^T\|_FK_{ij_1}K_{ij_2}=O_p(1),\\
    &\frac{h^2}{n}\sum^n_{i=1}\sum_{j_1=1}^m\sum_{j_2=1}^m|X_{ik}(\tau_{ij_1})X_{il}(\tau_{ij_2})\varepsilon_i(\tau_{ij_2})|\|X_i(\tau_{ij_1})\|_2K_{ij_1}K_{ij_2}=O_p(1), \mbox{ and}\\
    &\frac{h^2}{n}\sum^n_{i=1}\sum_{j_1=1}^m\sum_{j_2=1}^m1(\tau_{ij_1}\vee \tau_{ij_2}\leq T_i\le C_i)X_{ik}(\tau_{ij_1})X_{il}(\tau_{ij_2})\varepsilon_i(\tau_{ij_1})\varepsilon_i(\tau_{ij_2})K_{ij_1}K_{ij_2}\nonumber\\
    &\quad\to_p\sum_{j=1}^{m}\eta(t_0,s_0;j,k,l)f_{\tau_j,T-\tau_j}(t_0,s_0)\sigma^2(t_0)\mu_0
\end{align*}
under Conditions \ref{cond:X8_bounded}, \ref{cond:X83_bounded} and \ref{cond:sigma4_bounded}, here $\|\cdot\|_F$ denotes the Frobenious norm. Thus it suffices to show
\begin{align}
    \sup_{(t,s)\in\mathcal{S}_0(h)}\left\|\boldsymbol{\beta}(t,s;h)-\widehat{\boldsymbol{\beta}}(t,s;h)\right\|_2\to_p0,
    \label{eq:beta_unif_consistency}
\end{align}
where 
$$\mathcal{S}_0(h) = \left\{ (t,s): \left( \frac{t-t_0}{h}, \frac{s-s_0}{h} \right) \in \mbox{supp} \, K  \right\}.$$

We divide the proof of (\ref{eq:beta_unif_consistency}) into four parts following Lemma \ref{lem:four_parts}:
\begin{align}
    &\sup_{(t,s)\in\mathcal{S}_0(h)}|A_n(t,s;h)_{kl}-A(t,s)_{kl}|\to_p0,\label{eq:4_1}\\
    &\sup_{(t,s)\in\mathcal{S}_0(h)}\|B_n(t,s;h)-B(t,s)\|_2\to_p0,\label{eq:4_2}\\
    &\lim\inf_n\inf_{(t,s)\in\mathcal{S}_0(h)}\lambda_1(A(t,s))>0,\label{eq:4_3}\\
    &\lim\sup_n\sup_{(t,s)\in\mathcal{S}_0(h)}\|B(t,s)\|_2<\infty.\label{eq:4_4}
\end{align}
In the above, (\ref{eq:4_1}) and (\ref{eq:4_2}) follow from Lemma \ref{lem:fg_euclidean} and results in \cite{nolan_u-processes_1987} and \cite{pollard_uniform_1995}; (\ref{eq:4_3}) follows from Condition \ref{cond:pd} and the continuity of $A(t,s)$; and (\ref{eq:4_4}) follows from the local boundedness of $B(t,s)$. We then obtain the consistency of the sandwich variance estimator. 

\section*{Acknowledgement}

The data reported here have been
supplied by the United States Renal Data System (USRDS). The interpretation
and reporting of these data are the responsibility of the authors
and in no way should be seen as an official policy or interpretation of the
U.S. government.

\bibliographystyle{biom}
\bibliography{mybib}
\end{document}